\documentclass[aps,prl,showpacs,superscriptaddress,twocolumn]{revtex4}
\usepackage{graphicx}
\usepackage{amssymb}
\begin{document}
\title{Spin dynamics in a superconductor / ferromagnet proximity system}
\author{C. Bell}
\affiliation{Magnetic and Superconducting Materials Group, Kamerlingh Onnes
Laboratorium, Universiteit Leiden, The Netherlands}
\author{S. Milikisyants}\affiliation{Molecular Nano-Optics and Spins Group,
Leiden Institute of Physics, Universiteit Leiden, The Netherlands}
\author{M. Huber}\affiliation{Molecular Nano-Optics and Spins Group, Leiden
Institute of Physics, Universiteit Leiden, The Netherlands}
\author{J. Aarts}
\affiliation{Magnetic and Superconducting Materials Group, Kamerlingh Onnes
Laboratorium, Universiteit Leiden, The Netherlands}
\begin{abstract}
The ferromagnetic resonance of thin sputtered Ni$_{80}$Fe$_{20}$ films grown on Nb is measured. By varying the
temperature and thickness of the Nb the role of the superconductivity on the whole ferromagnetic layer in these
heterostructures is explored. The change in the spin transport properties below the superconducting transition of
the Nb is found to manifest itself in the Ni$_{80}$Fe$_{20}$ layer by a sharpening in the resonance of the
ferromagnet, or a decrease in the effective Gilbert damping co-efficient. This dynamic proximity effect is in
contrast to low frequency studies in these systems, where the effect of the superconductor is confined to a small
region in the ferromagnet. We interpret this in terms of the spin pumping model.
\end{abstract}
 \maketitle

Most of the experiments in the field of superconductor (S) / ferromagnet (F) hybrids rely on measuring their
electrical transport characteristics. In S/F/S Josephson junctions the measured quantity is mainly the
supercurrent, which is for instance used to show the existence of $\pi$-junctions (where the phase of the order
parameter undergoes a change of phase by $\pi$) \cite{buzdin2005}. More recently, experiments involving a
half-metallic ferromagnet found the supercurrent in that case to be long-ranged, possibly due to the occurrence of
spin-triplet superconductivity \cite{keizer}. Also in F/S/F structures, the injection of spins in superconductors
is mostly measured and analyzed by following the changes in electrical resistance of the device \cite{guprb}. Both
for questions involving spin transport and for studying the nature of the superconducting correlations inside the
ferromagnet, it would be advantageous to avail of a method which measures changes of the F-layer properties as a
consequence of the superconductivity. In hybrids of normal metals (N) and ferromagnets similar questions are
currently addressed by ferromagnetic resonance (FMR) experiments in the microwave regime, which study the dynamic
behavior of the precessing ferromagnetic spin of the F-layer in good electrical contact with an N-layer. The decay
of the precessing magnetization {\bf m}, and therefore the power absorption, can be written in terms of the
Landau-Lifshitz-Gilbert equation as:
\begin{equation}
\partial_t {\bf m} = -\gamma {\bf m} \times {\bf H}_{eff} + \alpha \; {\bf m}
\times \partial_t {\bf m},
\end{equation}
where $\gamma$ is the gyromagnetic ratio, ${\bf H}_{eff}$ an effective magnetic field and $\alpha$ the Gilbert
constant which controls the damping. This parameter is often parametrized as $G = \alpha \gamma M_s$. In the F/N
case, the damping is caused in some part by the emission of spin polarized electrons in a direction perpendicular
to the interface, which leads to the spin pumping or spin battery effect. Hence the properties of the nearby
metals in a heterostructure play a critical role in the determining the FMR lineshape. This has been reviewed in
detail recently by Tserkovnyak {\it et al.} \cite{tserkovnyakRMP}. Few FMR experiments have been reported as yet
for F/S systems. A study was made on bulk RuSr$_2$GdCu$_2$O$_8$~\cite{fainstein}, but here superconductivity and
ferromagnetism are intrinsically mixed. On the other hand, there are by now a number of theoretical predictions
that have not been experimentally investigated \cite{brataasPRL2004, braude, shi,nussinov, zhao}. In this Letter
we address the proximity effect of a superconductor (Nb) on the FMR behavior of a strong ferromagnet
(Ni$_{80}$Fe$_{20}$, Permalloy, Py) and find significant changes in the lineshape, implying that the entire
magnetic layer is affected rather than the small distance of the superconducting coherence length in the
ferromagnet.

Our samples are grown on 0.5 mm thick Suprasil$^{\circledR}$2 quartz (lateral dimensions $\sim$ 3mm $\times$ 5mm)
by d$.$c$.$ magnetron sputtering at room temperature, in a vacuum system with a base pressure $<$
2$\times$10$^{-9}$ mbar. Deposition rates were $\sim 0.12$ nm/s for the Nb and $\sim 0.14$ nm/s for the Py, as
calibrated from low angle x-ray reflectivity. First we focus on three different samples. Sample A consists of
$q$/Nb(70)/Py(5), where $q$ stands for the quartz substrate and the numerals give the layer thickness in nm;
sample B is $q$/Nb(9)/Py(5), and sample C is $q$/Nb(70)/Py(2)/Nb(5). The critical temperature T$_C$ $\sim$ 8.2~K
of sample A was measured in an in-plane magnetic field of $\mu_0$H = 100 mT. The transition width was $<$ 30 mK,
and the resistance ratio R(300K)/R(10K) = 3.07. For sample B the 9 nm thick Nb layer is below the critical
thickness of such a polycrystalline sample, and does not superconduct in the measurement range of temperatures
presented. Sample B thus serves as a reference with similar interface characteristics, but no superconductivity.
Both samples have an unprotected F-layer; sample C has a thinner Py layer as well as a non-superconducting (Nb)
protection cap.

Due to stray magnetic fields in the sputtering chamber, the Py possesses an in-plane uniaxial induced anisotropy,
giving a coercive field $\mu_0$H$_C^{easy}$ = 3.5 mT for sample A (T = 10 K), with $\mathrm{H}_C^{\mathrm{easy}}
\! \sim 0.8 \times \mathrm{H}_C^{\mathrm{hard}}$. All data presented are measured with the applied field nominally
along the easy axis. The FMR measurements were made in a Bruker ElexSys E680 X-band electron paramagnetic
resonance (EPR) system operating at 9.5 GHz, equipped with a rectangular TE$_{102}$ cavity and a liquid helium
continuous flow cryostat. The input power was nominally 220 mW, attenuated before the resonance cavity by a factor
of $10^4$ (40 dB). The d$.$c$.$ magnetic field was applied with a modulation of $\mu_0$H = 0.5 mT at 100 kHz. The
samples were secured with teflon tape onto a quartz rod mounted vertically on a goniometer to allow control of the
film normal direction with respect to the applied field: this was optimized at room temperature to be
$90^{\circ}\! \! \pm 2^{\circ}$ by minimizing the center field of the FMR.
\begin{figure}[h]
\includegraphics[width=8.6cm]{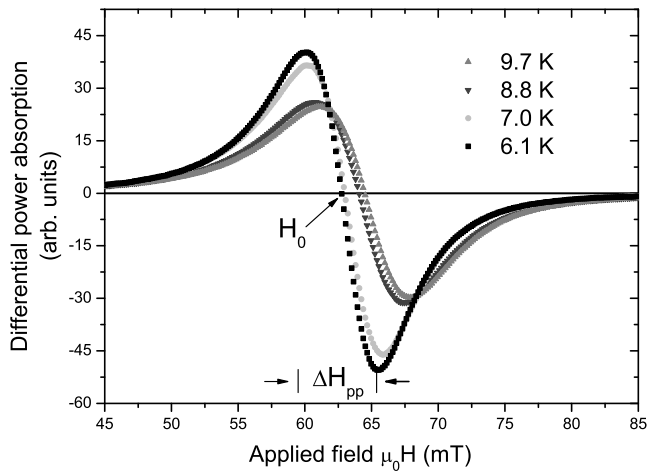}
\caption{Detail around the FMR of the EPR spectra for sample A taken around T$_C$. For clarity only half of the data
points for each curve are shown. $\Delta\mathrm{H}_{pp}$ and H$_0$ for the T = 6.1 K spectrum are labelled.}
\label{bellfigure1}
\end{figure}
Some typical EPR spectra taken for sample A are shown in Fig$.$ \ref{bellfigure1} above and below the
superconducting transition. The lineshape is close to the derivative of a Lorentzian line, with the ratio of the
amplitudes of the lobes above and below the baseline around 0.8. This asymmetry has been observed in many other
systems \cite{chappert}, and is associated with the polycrystalline nature of the samples, and variations of
saturation magnetization and anisotropy fields over the sample. In the figure we also define the two parameters
used to quantify changes in the resonance conditions. They are the zero-crossing field H$_0$ and the linewidth
$\Delta\mathrm{H}_{pp}$ (the field separation between the peak and dip position).

\begin{figure}[h]
\includegraphics[width=8.6cm]{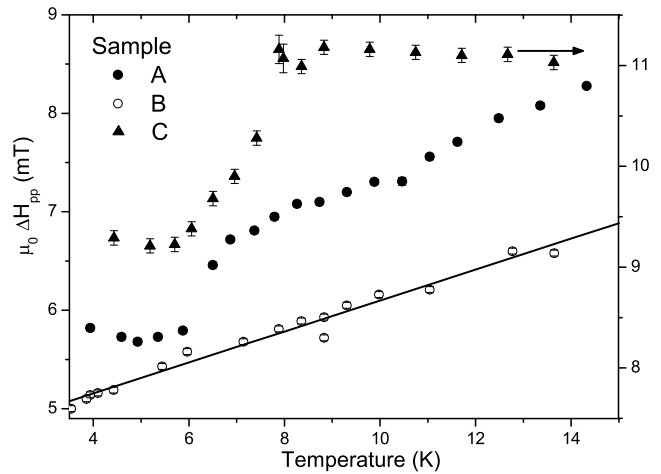}
\caption{$\Delta\mathrm{H}_{pp}$(T) data for samples A, B and C with T $<$ 15 K. Line is a guide to the eye.}
\label{bellfigure2}
\end{figure}

Fig$.$ \ref{bellfigure2} shows the central result of our paper. Here we plot $\Delta\mathrm{H}_{pp}$ versus
temperature T around T$_C$ for all three samples. The first thing to note is that the superconducting samples
(A,C) both show a significant non-monotonous decrease in linewidth when cooling through T$_C$, while the
non-superconducting sample (B) does not show such an anomaly. This indicates a strong {\it decrease} of the
damping experienced by the precessing magnetization in the F-layer when the adjacent S-layer becomes
superconducting. Before discussing this further we comment first on some other details of the data.

\begin{figure}[h]
\includegraphics[width=8.6cm]{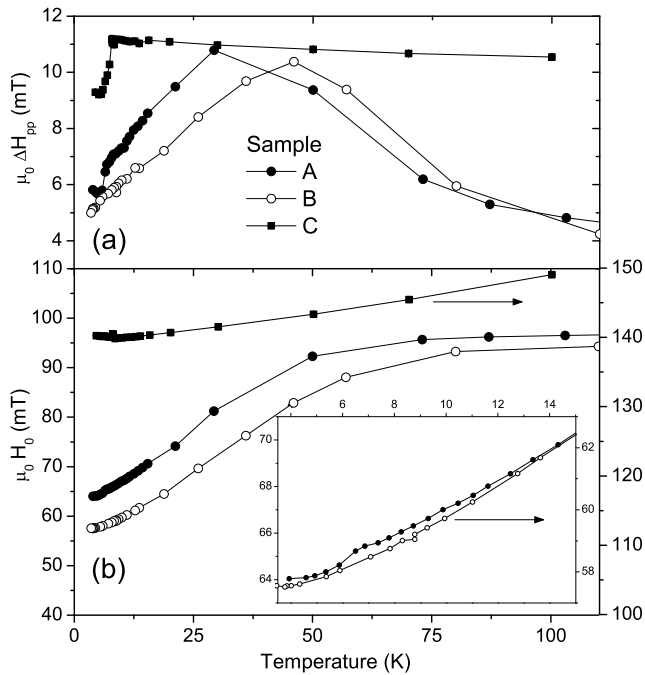}
\caption{(a)$\Delta\mathrm{H}_{pp}$ and (b) H$_0$ at higher T for samples A, B and C. Lines are a guide. Inset:
H$_0$ data for A and B below 15 K.} \label{bellfigure3}
\end{figure}

The values of $\Delta\mathrm{H}_{pp}$ in the normal state are different for all samples. The larger linewidth in
sample C is simply due to the much smaller thickness of the F-layer; the different values for samples A and B are
caused, we believe, by the thicker Nb-layer present in sample A and the influence of that layer on the
spin-pumping effect, as will be discussed below. Furthermore, still in the normal state,
$\Delta\mathrm{H}_{pp}$ for samples A,B show a clear temperature dependence which is absent in sample C.
This is caused by the difference between the capped and uncapped Py. To illustrate this point more clearly we show
the temperature dependence over a wider range in Fig$.$ \ref{bellfigure3}(a). For the uncapped samples,
$\Delta\mathrm{H}_{pp}$ actually goes through a maximum around 40 K, while the linewidth of the uncapped sample is
only weakly temperature dependent. This is in agreement with earlier work on single Py films, where similar
variations were found for films with a native oxide or with a magnetic (NiO) cap, while no temperature dependence
was observed for Cu-capped Py \cite{lubitz1998}. The native oxide apparently acts as a different magnetic system
which influences the Py, and this can also be seen in the temperature dependence of H$_0$, Fig$.$
\ref{bellfigure3}(b). For the uncapped samples H$_0$ shows a decrease below 50 K, while H$_0$ for sample C is
again changing only slightly in this regime. Around T$_C$ however, no anomalies are found in the behavior of H$_0$
(inset of Fig$.$ \ref{bellfigure3}(b)); the variations in $\Delta\mathrm{H}_{pp}$ are thus not due to variations
in the effective field experienced by the samples.

Returning to our main observation of the change in FMR linewidth at T$_C$, we wish to argue that this is due to
suppression of the spin-sinking mechanism which is provided by the normal layer, when this layer becomes
superconducting. For this, we first refer to a set of studies by Mizukami {\it et al.} \cite{mizukami_all} who
showed that the Gilbert damping coefficient $G$ slightly increased when a Cu layer of increasing thickness was
deposited on a thin Py film; whereas the linewidth became much larger when Pt was deposited instead; and that this
width decreased again when a Cu layer was inserted between the Py and the Pt. The basic explanation for these
observations (see \cite{tserkovnyakRMP}) is that the precessing moment drives spins into the normal metal, which
would be lost from the system if this is a clean and thick conductor; when the metal is dirty, however, spin flip
processes occur, making backscattering of electrons with opposite spin possible. This in turn leads to enhanced
damping. In other words, when the normal metal is a good spin sink, the damping will increase. This explains both
the small increase upon Cu deposition, and the much larger increase with a Pt layer. In the latter case, the
strong spin-orbit interaction enhances the spin-flip processes close to the interface. Needless to say, when a Cu
layer is placed between the Py and the Pt, the damping will decrease again.

In this model for spin pumping, a change in the Gilbert damping as a result of the normal metal-to-superconducting
transition can be understood since in the superconductor, spin transport is reduced. Electrons ejected from the
ferromagnet have energies well below the superconducting gap of Nb (1.5 meV at T = 0 K) and therefore cannot enter
in the superconductor as quasiparticles. The other mechanism to enter is through Andreev reflections, in which a
spin-up electron and a retro-reflected spin-down hole combine into a Cooper pair. This mechanism is partially
suppressed, however, since the spin subbands in the ferromagnet are not equally populated. Electrons therefore
cannot be emitted as efficiently as in the normal state, spin accumulation occurs at the S/F interface
\cite{jedemaPRB, jong}, and the amount of backscattering of electrons with opposite spin is reduced. The ensuing
decreased damping leads to the observed smaller linewidth. A similar change in spin transport has also been found
in d$.$c$.$ magnetoresistance measurements of Py/Nb/Py structures: below the superconducting transition the
effective spin diffusion length in the Nb decreases from around 50 nm to about 20 nm, close to the coherence
length $\xi_S$. This is to be expected since it is over this range that electrons are converted into Cooper pairs
\cite{guprb, yamashita2003}.

In our data we can now explain the difference between $\Delta\mathrm{H}_{pp}$ for samples A and B in the normal
state: the thicker Nb layer of sample A is a slightly more effective in backscattering opposite spins. This is
still also the case below T$_C$, which is fully consistent with a picture of a 20 nm spin diffusion length in the
Nb. One drawback to using Nb as the S layer is the relatively poor spin sink that it provides in the normal state.
In the limit of the spin sink thickness $d \gg \ell_{sd}$ the parameter $\epsilon=\frac{1}{3}(\ell/\ell_{sf})^2$
for Nb can be calculated using typical values for electron mean free path $\ell$ and spin flip length
$\ell_{sf}$\cite{guprb} giving $\epsilon=0.005$. This value is low compared to $\epsilon \ge 0.01$ which is
required for efficient spin sinking \cite{tserkovnyakRMP}. We must also consider the possible effects of the
presence of in-plane vortices in the superconductor. At 9.5 GHz vortices will in general not depin, but locally
oscillate in the local minima of the pinning potential. Any vortex motion would be expected to absorb microwaves
over a relatively broad field range, and thus broaden the resonance which is not observed in the present data.
\begin{figure}[h]
\includegraphics[width=8.6cm]{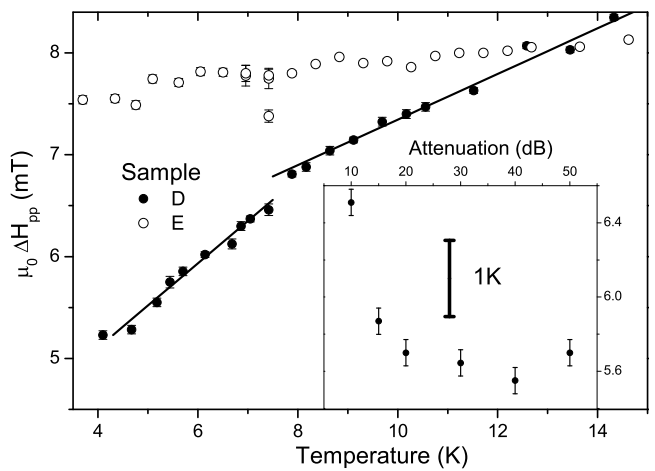}
\caption{$\Delta\mathrm{H}_{pp}$(T) for samples D and E. Lines are best fits above and below 7.7 K. Inset:
$\Delta\mathrm{H}_{pp}$ versus input power attenuation at T = 5.2 K for sample D. The heavy scale bar represents
the equivalent shift in $\Delta\mathrm{H}_{pp}$ for a 1 K temperature change from the best fit line below 7.7 K.}
\label{bellfigure4}
\end{figure}

We have argued that it is variations in spin sinking efficiency which leads to the changes in linewidths when
passing through $T_C$. We finish by showing that, just as in normal metals, the effects of added spin scattering
by spin-orbit coupling can be utilized for further studies. To this end we fabricated two more samples. Sample D
consists of $q$/Pt(20)/Nb(70)/Py(5nm) with a T$_C$ = 8.0 K. The Py layer is uncapped while the Nb layer is thick,
with Pt as an underlayer. Sample E is $q$/NbN(50)/Pt(5)/Nb(25)/Py(5), with T$_C$ = 10.6 K. The Py layer is also
uncapped while here the Pt layer is in contact with a much thinner Nb layer. Since this considerably lowers the
superconducting transition we then boost the superconductivity again by a NbN layer. The extracted values for
$\Delta\mathrm{H}_{pp}$ for these samples are shown in Fig$.$ \ref{bellfigure4}. The $\Delta\mathrm{H}_{pp}$(T)
for sample D is essentially the same as the data for sample A (although no clear saturation is observed at the
lowest temperatures). This weak effect of Pt is caused by the relatively thick Nb layer, meaning that most of the
spin memory is in the Nb before reaching the Pt. For sample E, however, the Pt sits no more than the (Nb) spin
diffusion length away from the Py-layer, spin sinking is quite efficient, and the T-dependence of
$\Delta\mathrm{H}_{pp}$ is suppressed even below T$_C$. We note that the EPR spectra for sample D were
significantly more noisy than the other samples, for reasons which are not clear at this time. Finally we show in
the inset of Fig$.$ \ref{bellfigure4} that there is no direct heating of the films by the microwaves at low T
until an attenuation of $10-15$ dB, which is greater than the power used for our measurements.

To summarize, we have shown that a superconductor in good metallic contact with a ferromagnet can influence the
dynamics of the whole ferromagnetic layer by decreasing the spin sink efficiency, leading to a decreased Gilbert
damping of the FMR. This is consistent with a picture of suppressed Andreev reflections at the S/F interface due
to the spin polarization of the Py causing a reduced spin transport. Although currently there is no theoretical
description of this system, a full theoretical treatment is forthcoming \cite{morten}. The decrease in the damping
of the F layer is highly relevant for S/F hybrid devices with strong ferromagnets
\cite{blum,bellapl,bellPRB2005,robinson,born} at high frequencies, in for example possible coupling between the
FMR and the a$.$c$.$ Josephson effect \cite{WaintalPRB}, as well as enhanced magnetic noise in devices utilizing
S/F/S junctions. These issues have not been addressed so far. We would like to stress the potential usefulness of
the EPR technique in the S/F field, as well as for research into spin pumping. These S/F systems can be used
beneficially to access a range of spin transport properties in a single system, foregoing the need for many
samples where variations in film properties can give rise to non-intrinsic contributions \cite{mizukami_all}.

Although the effects presented here can be satisfactorily explained by the superconducting gap opening in the Nb
layer, an interesting question is if any additional effects can be observed due to the inhomogeneous induced
superconducting state in the Py. Recent measurements indicate a superconducting coherence length in Py of $1- 1.5$
nm \cite{bellPRB2005,robinson}, meaning that a significant fraction of the Py in this experiment has some
superconductivity induced in it. It is also therefore interesting to measure the FMR in weak ferromagnets where
the induced superconductivity is relatively long range \cite{ryazanov1, kontos}. In this case the lower spin
polarization of the F layers will lead to less spin accumulation at the S/F interface, hence a weaker suppression
of the spin pumping, and thus a relatively large contribution from any novel effects. Further studies with F/S/F
trilayers with non-collinear F layers or a S/half metal system would also be especially interesting since any
triplet superconducting components \cite{fominov,keizer} are expected contribute to the FMR damping in a different
way to the singlet component \cite{brataasPRL2004}.

We are very happy to thank J. Schmidt, G.E.W. Bauer, A. Brataas and J.P. Morten for stimulating discussions and
suggestions, and R.W.A. Hendrikx for x-ray measurements. This work was supported by the Dutch ``Stichting FOM''.


\begin{thebibliography}{26}
\expandafter\ifx\csname natexlab\endcsname\relax\def\natexlab#1{#1}\fi
\expandafter\ifx\csname bibnamefont\endcsname\relax
  \def\bibnamefont#1{#1}\fi
\expandafter\ifx\csname bibfnamefont\endcsname\relax
  \def\bibfnamefont#1{#1}\fi
\expandafter\ifx\csname citenamefont\endcsname\relax
  \def\citenamefont#1{#1}\fi
\expandafter\ifx\csname url\endcsname\relax
  \def\url#1{\texttt{#1}}\fi
\expandafter\ifx\csname urlprefix\endcsname\relax\def\urlprefix{URL }\fi
\providecommand{\bibinfo}[2]{#2}
\providecommand{\eprint}[2][]{\url{#2}}

\bibitem[{\citenamefont{Buzdin}(2005)}]{buzdin2005}
\bibinfo{author}{\bibfnamefont{A.~I.} \bibnamefont{Buzdin}},
  \bibinfo{journal}{Rev. Mod. Phys.} \textbf{\bibinfo{volume}{77}},
  \bibinfo{pages}{935} (\bibinfo{year}{2005}).

\bibitem[{\citenamefont{{R. S. Keizer {\it et al.}}}(2006)}]{keizer}
\bibinfo{author}{\bibnamefont{{R. S. Keizer {\it et al.}}}},
  \bibinfo{journal}{Nature} \textbf{\bibinfo{volume}{439}},
  \bibinfo{pages}{825} (\bibinfo{year}{2006}).

\bibitem[{\citenamefont{{J. Y. Gu {\it et al.}}}(2002)}]{guprb}
\bibinfo{author}{\bibnamefont{{J. Y. Gu {\it et al.}}}},
  \bibinfo{journal}{Phys. Rev. B} \textbf{\bibinfo{volume}{66}},
  \bibinfo{pages}{140507} (\bibinfo{year}{2002}).

\bibitem[{\citenamefont{{Y. Tserkovnyak {\it et al.}}}(2005)}]{tserkovnyakRMP}
\bibinfo{author}{\bibnamefont{{Y. Tserkovnyak {\it et al.}}}},
  \bibinfo{journal}{Rev. Mod. Phys.} \textbf{\bibinfo{volume}{77}},
  \bibinfo{pages}{1375} (\bibinfo{year}{2005}).

\bibitem[{\citenamefont{{A. Fainstein {\it et al.}}}(1999)}]{fainstein}
\bibinfo{author}{\bibnamefont{{A. Fainstein {\it et al.}}}},
  \bibinfo{journal}{Phys. Rev. B} \textbf{\bibinfo{volume}{60}},
  \bibinfo{pages}{R12597} (\bibinfo{year}{1999}).

\bibitem[{\citenamefont{Brataas and Tserkovnyak}(2004)}]{brataasPRL2004}
\bibinfo{author}{\bibfnamefont{A.}~\bibnamefont{Brataas}} \bibnamefont{and}
  \bibinfo{author}{\bibfnamefont{Y.}~\bibnamefont{Tserkovnyak}},
  \bibinfo{journal}{Phys. Rev. Lett.} \textbf{\bibinfo{volume}{93}},
  \bibinfo{pages}{087201} (\bibinfo{year}{2004}).

\bibitem[{\citenamefont{Braude}(2006)}]{braude}
\bibinfo{author}{\bibfnamefont{V.}~\bibnamefont{Braude}},
  \bibinfo{journal}{Phys. Rev. B} \textbf{\bibinfo{volume}{74}},
  \bibinfo{pages}{054515} (\bibinfo{year}{2006}).

\bibitem[{\citenamefont{{Y.-M. Shi {\it et al.}}}(2006)}]{shi}
\bibinfo{author}{\bibnamefont{{Y.-M. Shi {\it et al.}}}},
  \bibinfo{journal}{Europhys. Lett.} \textbf{\bibinfo{volume}{73}},
  \bibinfo{pages}{941} (\bibinfo{year}{2006}).

\bibitem[{\citenamefont{{Z. Nussinov {\it et al.}}}(2005)}]{nussinov}
\bibinfo{author}{\bibnamefont{{Z. Nussinov {\it et al.}}}},
  \bibinfo{journal}{Phys. Rev. B} \textbf{\bibinfo{volume}{71}},
  \bibinfo{pages}{214520} (\bibinfo{year}{2005}).

\bibitem[{\citenamefont{Zhao et~al.}(2004)\citenamefont{Zhao, L{\"o}fwander,
  and Sauls}}]{zhao}
\bibinfo{author}{\bibfnamefont{E.}~\bibnamefont{Zhao}},
  \bibinfo{author}{\bibfnamefont{T.}~\bibnamefont{L{\"o}fwander}},
  \bibnamefont{and} \bibinfo{author}{\bibfnamefont{J.~A.} \bibnamefont{Sauls}},
  \bibinfo{journal}{Phys. Rev. B} \textbf{\bibinfo{volume}{70}},
  \bibinfo{pages}{134510} (\bibinfo{year}{2004}).

\bibitem[{\citenamefont{{C. Chappert {\it et al.}}}(1986)}]{chappert}
\bibinfo{author}{\bibnamefont{{C. Chappert {\it et al.}}}},
  \bibinfo{journal}{Phys. Rev. B} \textbf{\bibinfo{volume}{34}},
  \bibinfo{pages}{3192} (\bibinfo{year}{1986}).

\bibitem[{\citenamefont{{P. Lubitz {\it et al.}}}(1998)}]{lubitz1998}
\bibinfo{author}{\bibnamefont{{P. Lubitz {\it et al.}}}}, \bibinfo{journal}{J.
  Appl. Phys.} \textbf{\bibinfo{volume}{83}}, \bibinfo{pages}{6819}
  (\bibinfo{year}{1998}).

\bibitem[{\citenamefont{Mizukami et~al.}()\citenamefont{Mizukami, Ando, and
  Miyazaki}}]{mizukami_all}
\bibinfo{author}{\bibfnamefont{S.}~\bibnamefont{Mizukami}},
  \bibinfo{author}{\bibfnamefont{Y.}~\bibnamefont{Ando}}, \bibnamefont{and}
  \bibinfo{author}{\bibfnamefont{T.}~\bibnamefont{Miyazaki}},
  \bibinfo{title}{{Phys. Rev. B {\bf 66}, 104413 (2002); J. Magn. Magn.
  Mater. {\bf 226-230}, 1640 (2001); Jpn. J. Appl. Phys. {\bf 40}, 580
  (2001)}}.

\bibitem[{\citenamefont{{F. J. Jedema {\it et al.}}}(1999)}]{jedemaPRB}
\bibinfo{author}{\bibnamefont{{F. J. Jedema {\it et al.}}}},
  \bibinfo{journal}{Phys. Rev. B} \textbf{\bibinfo{volume}{60}},
  \bibinfo{pages}{16549} (\bibinfo{year}{1999}).

\bibitem[{\citenamefont{de~{J}ong and Beenakker}(1995)}]{jong}
\bibinfo{author}{\bibfnamefont{M.~J.~M.} \bibnamefont{de~{J}ong}}
  \bibnamefont{and} \bibinfo{author}{\bibfnamefont{C.~W.~J.}
  \bibnamefont{Beenakker}}, \bibinfo{journal}{Phys. Rev. Lett.}
  \textbf{\bibinfo{volume}{74}}, \bibinfo{pages}{1657} (\bibinfo{year}{1995}).

\bibitem[{\citenamefont{{T. Yamashita {\it et al.}}}(2003)}]{yamashita2003}
\bibinfo{author}{\bibnamefont{{T. Yamashita {\it et al.}}}},
  \bibinfo{journal}{Phys. Rev. B} \textbf{\bibinfo{volume}{67}},
  \bibinfo{pages}{094515} (\bibinfo{year}{2003}).

\bibitem[{\citenamefont{{J. P. Morten {\it et al.}}}()}]{morten}
\bibinfo{author}{\bibnamefont{{J. P. Morten {\it et al.}}}},
  \bibinfo{howpublished}{{(private communication)}}.

\bibitem[{\citenamefont{{Y. Blum {\it et al.}}}(2002)}]{blum}
\bibinfo{author}{\bibnamefont{{Y. Blum {\it et al.}}}}, \bibinfo{journal}{Phys.
  Rev. Lett.} \textbf{\bibinfo{volume}{89}}, \bibinfo{pages}{187004}
  (\bibinfo{year}{2002}).

\bibitem[{\citenamefont{{C. Bell {\it et al.}}}(2004)}]{bellapl}
\bibinfo{author}{\bibnamefont{{C. Bell {\it et al.}}}}, \bibinfo{journal}{Appl.
  Phys. Lett.} \textbf{\bibinfo{volume}{84}}, \bibinfo{pages}{1153}
  (\bibinfo{year}{2004}).

\bibitem[{\citenamefont{{C. Bell {\it et al.}}}(2005)}]{bellPRB2005}
\bibinfo{author}{\bibnamefont{{C. Bell {\it et al.}}}}, \bibinfo{journal}{Phys.
  Rev. B} \textbf{\bibinfo{volume}{71}}, \bibinfo{pages}{180501(R)}
  (\bibinfo{year}{2005}).

\bibitem[{\citenamefont{{J. W. A. Robinson {\it et al.}}}(2006)}]{robinson}
\bibinfo{author}{\bibnamefont{{J. W. A. Robinson {\it et al.}}}},
  \bibinfo{journal}{Phys. Rev. Lett.} \textbf{\bibinfo{volume}{97}},
  \bibinfo{pages}{177003} (\bibinfo{year}{2006}).

\bibitem[{\citenamefont{{F. Born {\it et al.}}}(2006)}]{born}
\bibinfo{author}{\bibnamefont{{F. Born {\it et al.}}}}, \bibinfo{journal}{Phys.
  Rev. B} \textbf{\bibinfo{volume}{74}}, \bibinfo{pages}{{140501(R)}}
  (\bibinfo{year}{2006}).

\bibitem[{\citenamefont{Waintal and Brouwer}(2002)}]{WaintalPRB}
\bibinfo{author}{\bibfnamefont{X.}~\bibnamefont{Waintal}} \bibnamefont{and}
  \bibinfo{author}{\bibfnamefont{P.~W.} \bibnamefont{Brouwer}},
  \bibinfo{journal}{Phys. Rev. B} \textbf{\bibinfo{volume}{65}},
  \bibinfo{pages}{054407} (\bibinfo{year}{2002}).

\bibitem[{\citenamefont{{V. V. Ryazanov {\it et al.}}}(2001)}]{ryazanov1}
\bibinfo{author}{\bibnamefont{{V. V. Ryazanov {\it et al.}}}},
  \bibinfo{journal}{Phys. Rev. Lett.} \textbf{\bibinfo{volume}{86}},
  \bibinfo{pages}{2427} (\bibinfo{year}{2001}).

\bibitem[{\citenamefont{{T. Kontos {\it et al.}}}(2001)}]{kontos}
\bibinfo{author}{\bibnamefont{{T. Kontos {\it et al.}}}},
  \bibinfo{journal}{Phys. Rev. Lett.} \textbf{\bibinfo{volume}{86}},
  \bibinfo{pages}{304} (\bibinfo{year}{2001}).

\bibitem[{\citenamefont{Fominov et~al.}(2003)\citenamefont{Fominov, Golubov,
  and Kupriyanov}}]{fominov}
\bibinfo{author}{\bibfnamefont{Y.~V.} \bibnamefont{Fominov}},
  \bibinfo{author}{\bibfnamefont{A.~A.} \bibnamefont{Golubov}},
  \bibnamefont{and} \bibinfo{author}{\bibfnamefont{M.~Y.}
  \bibnamefont{Kupriyanov}}, \bibinfo{journal}{JETP Lett.}
  \textbf{\bibinfo{volume}{77}}, \bibinfo{pages}{510} (\bibinfo{year}{2003}).

\end{thebibliography}
\end{document}